\documentclass[pra,aps,twocolumn,eqsecnum,showpacs]{revtex4}

\usepackage{dcolumn}
\usepackage{amsmath}
\usepackage{graphics}

\begin{document}

\title{Spectral theory of quantum memory and entanglement \\via Raman
scattering of light by an atomic ensemble}
\author{O.S. Mishina, D.V. Kupriyanov}
\affiliation{Department of Theoretical Physics, State Polytechnic
University, 195251, St.-Petersburg, Russia}%
\email{Kupr@DK11578.spb.edu}%
\author{J.H. M\"{u}ller, E.S. Polzik}
\affiliation{QUANTOP - Danish Quantum Optics Center, Niels Bohr
Institute, 2100 Copenhagen, Denmark}%
\email{polzik@nbi.dk}%

\date{\today }

\begin{abstract}
We discuss theoretically quantum interface between light and a
spin polarized ensemble of atoms with the spin $\geq 1$ based on
an off-resonant Raman scattering. We present the spectral theory
of the light-atoms interaction and show how particular spectral
modes of quantum light couple to spatial modes of the extended
atomic ensemble. We show how this interaction can be used for
quantum memory storage and retrieval and for deterministic
entanglement protocols. The proposed protocols are attractive due
to their simplicity since they involve just a single pass of light
through atoms without the need for elaborate pulse shaping or
quantum feedback. As a practically relevant example we consider
the interaction of a light pulse with hyperfine components of
$D_1$ line of ${}^{87}$Rb. The quality of the proposed protocols
is verified via analytical and numerical analysis.
\end{abstract}

\pacs{03.67.Mn, 34.50.Rk, 34.80.Qb, 42.50.Ct}%
\maketitle%


\section{Introduction}
Light-matter quantum interface is a basic element of any quantum
information network aiming at long distance quantum communication,
cryptography protocols or quantum computing, see
\cite{SaBPtVL,Ralph}. Light is a natural carrier of quantum
information and a macroscopic atomic system can be efficiently
used for its storage. Critical ingredients of the quantum
interface are the quantum memory and entanglement protocols, which
allow high-fidelity interchange (transfer, storage and readout) of
quantum states between the light and relatively long-lived atomic
subsystems. A number of promising theoretical proposals for the
high-fidelity memory and entanglement protocols has been put
forward, which can be classified as based on off-resonant
interaction, such as Raman interaction \cite{KMP}, quantum
non-demolition (QND) interaction with quantum feedback - QNDF
\cite{QNDF}, multiple QND interactions Ref.\cite{MHPC}, and on
resonant interaction using electromagnetically induced
transparency - EIT \cite{FlLk,GAFSL}. In spite of a large number
of proposals, experimental realization of the complete storage
plus retrieval quantum memory with the fidelity higher than
classical is yet to be achieved. High fidelity storage (but not
complete retrieval) has been demonstrated via the QNDF approach
\cite{JSFCP} and as light-to-atoms teleportation \cite{SKOJHCP}.
Low fidelity storage and retrieval of the coherent and single
photon pulses based on EIT \cite{EIT} and Raman \cite{Kimble}
processes have been recently demonstrated.

All experiments on light-atoms quantum interface to date are
conducted with alkali atoms. Previously the off-resonant quantum
interface protocols \cite{JSFCP,SKOJHCP} utilized light with the
detuning higher than the hyperfine splitting of the excited state.
Under this condition magnetic sublevels of a hyperfine state of an
alkali atom can be effectively reduced to a spin-half system. As
shown in the present paper eliminating this restriction and thus
using the complete magnetic multipole system of the alkali atom
ground state opens up new possibilities for quantum memory and
entanglement generation with ensembles of such atoms. The quantum
description of correlations in coherent Raman scattering in the
Heisenberg formalism has a long history
\cite{RaymMost,DBP,WasRaym,NWRSWWJ}. However in quantum
information applications the Raman process is often discussed in
terms of formal $\Lambda$-scheme configuration. In contrast in our
treatment we describe this process in terms of polarization
multipoles for multilevel ground state alkali atoms: gyrotropy
(orientation), linear birefringence (alignment) and higher
multipole components. The field subsystem is described by a set of
polarization Stokes components. This approach allows for complete
description of the mean values and fluctuations of the light and
optically thick atomic medium. Although examples treated in this
paper concern atoms initially pumped to one level corresponding to
a single $\Lambda$-scheme (as shown in figures \ref{Fig.1}and
\ref{Fig.2}), our formalism also allows to treat more complex
initial states consisting of several coupled $\Lambda$-schemes.
The formalism is also applicable beyond the Heisenberg equations
in a linearized form. Finally, the multipole expansion facilitates
systematic and compact calculation of the coupling constants in
the effective Hamiltonian taking into account the full hyperfine
structure of the relevant atomic levels \cite{KMSJP}.

We shall discuss the Raman quantum memory scheme first considered
in \cite{KMP}. We shall add the retrieval step to the protocol and
analyze the complete procedure for a realistic model, taking full
account of the multilevel hyperfine and magnetic sublevel
structure of an alkali atom. We derive the polarization-sensitive
coupled wave-type dynamics of an optically thick atomic ensemble
and optical field. Using spectral mode decomposition for light and
spatial mode picture for the atomic ensemble we show that the
quantum states of light and the quasi-spin of atoms can be
effectively swapped or entangled. The general mathematical
formalism for such atomic system with angular momentum higher or
equal than one, has been developed in Refs.\cite{KMSJP,MKP}. An
important advantage of the memory and entanglement schemes in such
a scenario is in that they can be realized in a single pass of
light through atoms and without any feedback channel, which makes
experimental realizations more feasible.

As an elementary carrier of the quantum information we consider a
squeezed state of light. The relevance of quantum memory for these
states for quantum information processing has been illustrated in
proposals \cite{Preskill}.  We introduce Stokes operators for a
light pulse consisting of a circularly polarized strong classical
mode and the quantum squeezed light in the orthogonal circularly
polarized mode. We show that for the quantum Stokes variables
there is a convenient symmetric interaction with the alignment
tensor components of atoms, whose spin angular momenta are equal
or greater than one. In the quantum memory scenario the quantum
information of the light subsystem can be effectively mapped into
the alignment subsystem of atoms. In the entanglement scenario the
excitation of atomic spins with coherent pulse generates a
parametric-type interaction process, which results in creation of
the strong correlations between the quantum fluctuations of the
Stokes components of the transmitted light and the alignment
components in the spin subsystem of atoms.

We consider the interaction via the $D_1$ line of ${}^{87}$Rb, as
an important example, where a convenient spin-one system exists in
the lower hyperfine sublevel of its ground state. For verification
of the memory protocol we discuss its figure of merit for the
mapping of the input squeezed state and for its retrieval after
reading the quantum copy out from the spin subsystem with a second
coherent light pulse. We discuss how the fidelity for the proposed
quantum memory channel could be defined and compared with the
respective classical benchmark based on direct measurement of the
squeezed state parameters. We show that in an optimal
configuration the quantum fidelity is always higher than the limit
for the optimal competing classical channel.

Throughout our analysis we neglect coupling of the ground state
atomic degrees of freedom to other variables than the forward
propagating light field. This approximation is justified on time
scales which ar short compared to dephasing time of the ground
state coherence. Neglecting coupling to light modes propagating in
other than the forward direction can only be a good approximation
in the limit of high optical depth as detailed in section
\ref{IIIA}. Finally, the influence of the atomic motion capable of
washing out spatial spin patterns is not taken into account, which
restricts our analysis to cold atomic samples.

\section{Overview of the physical processes and basic assumptions}
\subsection{The schemes of experiments under discussion}
\begin{figure}[tp]
\vspace{\baselineskip}
\includegraphics{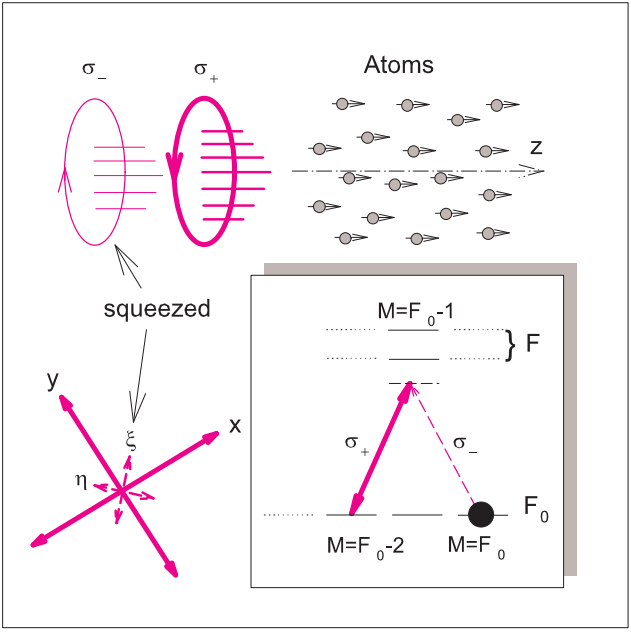}
\caption{Schematic diagram showing the geometry of the proposed
experiment on quantum memory and the scheme of relevant excitation
transitions, see text for details.}
\label{Fig.1}%
\end{figure}

\begin{figure}[tp]
\vspace{\baselineskip}
\includegraphics{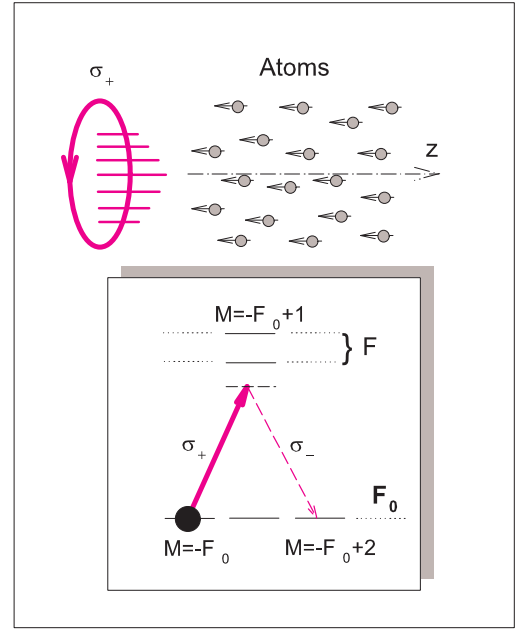}
\caption{Schematic diagram showing the geometry of the proposed
experiment on entanglement between light and atomic subsystems and
the scheme of relevant excitation transitions, see text for
details.}
\label{Fig.2}%
\end{figure}

In this paper we will consider two alternative experimental
situations respectively shown in figures \ref{Fig.1} and
\ref{Fig.2}. For both schemes a 100\% right-hand circular
polarized classical light pulse interacts with an ensemble of
ultracold spin-oriented atoms. The principal difference is in the
direction of the collective spin of the atomic ensemble, which can
be oriented either along (Fig.\ref{Fig.1}) or opposite
(Fig.\ref{Fig.2}) to the propagation direction of light.

Consider first the process shown in figure \ref{Fig.1}. For an
off-resonant right-hand polarized pulse, such that incoherent
scattering losses do not frustrate the spin polarization, there
will be no interaction between the light and atoms. However if a
small portion of a left-hand polarized quantum informative light
prepared in an unknown squeezed vacuum state is admixed to the
classical coherent pulse, the coherent scattering channel will be
open. The portion of the weak quantum light will be coherently
scattered into the strong classical mode. In this coherent Raman
process the polarization quantum subsystem of the probe light and
the spin subsystem of atoms can effectively swap their quantum
states. The quantum state can be mapped into the alignment-type
fluctuations of the spin subsystem and further stored in the form
of a certain standing spin wave for relatively long time. It can
be readout on demand with a second probe light pulse.

In the experimental situation shown in figure \ref{Fig.2}, for a
transparent medium a small portion of the left-hand polarized
photons will emerge as a result of the coherent elastic Raman
scattering of the strong light pulse. This process creates
entanglement the left-hand polarized quantum modes of light and
the alignment-type quantum fluctuations in the atomic spin
subsystem. This kind of deterministic entanglement can be
interesting for implementing a quantum repeater protocol between
remote atomic systems.

The processes shown in figures \ref{Fig.1} and \ref{Fig.2} have
practically identical theoretical description. Thus we will
discuss them in parallel paying attention to the differences in
their physical nature. We begin by introducing the relevant set of
the field and atomic variables involved in these processes.

\subsection{Field subsystem}

The polarization states of the light subsystem can be specified in
terms of the following flux-type variables for different Stokes
components. The Heisenberg operators for the total photon flux
considered at the spatial point $z$ and at the time moment $t$ are
given by
\begin{equation}
\hat{\Xi}_0(z,t)\;=\;%
\frac{S_0\,c}{2\pi\hbar\,\bar{\omega}}\,%
\hat{\mathbf{E}}^{(-)}(z,t)\,\hat{\mathbf{E}}^{(+)}(z,t)%
\label{2.1}%
\end{equation}
where $\hat{\mathbf{E}}^{(\pm)}(z,t)$ are the positive/negative
frequency components of the electric field Heiseberg operators. We
assume the quasi-monochromatic and forward propagating probe light
such that the spectral envelope of the modes has a carrier
frequency $\bar{\omega}$, and the light beam has a cross area
$S_0$.

The Stokes components responsible for three alternative
polarization types are given by
\begin{eqnarray}
\hat{\Xi}_1(z,t)&=&%
\frac{S_0\,c}{2\pi\hbar\,\bar{\omega}}\,%
\left[\hat{E}_\xi^{(-)}(z,t)\,\hat{E}_\xi^{(+)}(z,t)\right.%
\nonumber\\%
&&\phantom{\frac{S_0\,c}{2\pi\hbar\,\bar{\omega}}\,}%
\left.-\,\hat{E}_\eta^{(-)}(z,t)\,\hat{E}_\eta^{(+)}(z,t)\right]%
\nonumber\\%
\hat{\Xi}_2(z,t)&=&%
\frac{S_0\,c}{2\pi\hbar\,\bar{\omega}}\,%
\left[\hat{E}_R^{(-)}(z,t)\,\hat{E}_R^{(+)}(z,t)\right.%
\nonumber\\%
&&\phantom{\frac{S_0\,c}{2\pi\hbar\,\bar{\omega}}\,}%
\left.-\,\hat{E}_L^{(-)}(z,t)\,\hat{E}_L^{(+)}(z,t)\right]%
\nonumber\\%
\hat{\Xi}_3(z,t)&=&%
\frac{S_0\,c}{2\pi\hbar\,\bar{\omega}}\,%
\left[\hat{E}_x^{(-)}(z,t)\,\hat{E}_x^{(+)}(z,t)\right.%
\nonumber\\%
&&\phantom{\frac{S_0\,c}{2\pi\hbar\,\bar{\omega}}\,}%
\left.-\,\hat{E}_y^{(-)}(z,t)\,\hat{E}_y^{(+)}(z,t)\right]%
\label{2.2}%
\end{eqnarray}%
These components subsequently define imbalance in the photon
fluxes with respect to the $x/y$ Cartesian basis ($\hat{\Xi}_3$),
to the $\xi/\eta$ Cartesian basis rotated at $\pi/4$-angle
($\hat{\Xi}_1$), and to the $R/L$ basis of circular polarizations
($\hat{\Xi}_2$), see \cite{footnote}.

The Stokes variables (\ref{2.2}) obey the following commutation
relations
\begin{equation}
\left[\hat{\Xi}_{\mathrm i}(z,t), \hat{\Xi}_{\mathrm j}(z',t)\right]\;=\;%
2i\varepsilon_{\mathrm ijk}\,c\,\delta(z-z')\hat{\Xi}_k(z,t)%
\label{2.3}%
\end{equation}
where $\varepsilon_{\mathrm ijk}=\pm 1$ depending on the order of
the indices ${\mathrm i}\neq {\mathrm j}\neq {\mathrm k}$. The
$\delta$-function on the right-hand side of this commutation
relation is an approximation valid if only a finite bandwidth of
the field continuum is important for the correct description of
the low frequency fluctuation spectrum. This assumption is in
accord with the rotating wave approximation which we will further
use in the description of the atom-field interaction. For the
processes shown in figures \ref{Fig.1} and \ref{Fig.2} the
commutation relations (\ref{2.3}) can be simplified to
\begin{equation}
\left[\hat{\Xi}_3(z,t), \hat{\Xi}_1(z',t)\right]\;=\;%
2i\,c\,\delta(z-z')\bar{\Xi}_2%
\label{2.4}%
\end{equation}
where for the case of small fluctuations the operator on the
right-hand side is replaced by its expectation value. This Stokes
component is approximately conserved in the linear evolution, and
because of the conservation law for the number of photons in the
coherent process one has $\bar{\Xi}_2=\bar{\Xi}_0$. The component
$\bar{\Xi}_0$ is the integral of motion in our model and
$N_P=\bar{\Xi}_0T$, where $T$ is the interaction time, gives the
average number of photons participating in the process.

\subsection{Atomic spin subsystem}

The angular momentum polarization of $a$-th atom can be
conveniently described in the formalism of the irreducible tensor
operators
\begin{eqnarray}
\lefteqn{\hat{T}_{KQ}^{(a)}=\sqrt{\frac{2K+1}{2F_0+1}}\,%
\sum_{M',M}\,C_{F_0M\,KQ}^{F_0M'}\,|F_0M'\rangle\langle F_0M|^{(a)}}%
\nonumber \\%
&&|F_0M'\rangle\langle F_0M|^{(a)}=\sum_{KQ}\,%
\sqrt{\frac{2K+1}{2F_0+1}}\,%
C_{F_0M\,KQ}^{F_0M'}\,\hat{T}_{KQ}^{(a)}%
\nonumber\\%
\label{2.5}%
\end{eqnarray}%
With these expansions the set of atomic dyadic operators
originally defined in the subspace of the Zeeman states
$|F_0M\rangle$, where $F_0$ is the total (spin + nuclear) angular
momentum of atom and $M$ is its Zeeman projection, is transformed
into the set of the relevant irreducible tensor operators. Here
$C_{\ldots\,\ldots}^{\ldots}$ are the Clebsh-Gordan coefficients,
with the rank and the projection $K,Q$ which respectively vary in
the intervals: $0\leq K\leq 2F_0$ and $-K\leq Q\leq K$, see
Ref.\cite{VMK}.

The Heisenberg dynamics of the irreducible components (\ref{2.5})
preserves the following commutation relations
\begin{eqnarray}
\left[\hat{T}_{KQ}^{(a)}(t), \hat{T}_{K'Q'}^{(b)}(t)\right]&=&%
\delta_{ab}\,[(2K+1)(2K'+1)]^{1/2}\,%
\nonumber\\%
&&\hspace{-0.8in}\times \sum_{K''}\;[1-(-1)^{K+K'+K''}]\,%
\left\{\begin{array}{ccc}%
K & K' & K''\\ F_0 & F_0 & F_0%
\end{array}\right\}\,%
\nonumber\\%
&&\hspace{-0.8in}\times\,(-1)^{2F_0+K''}\,C^{K''Q''}_{KQ\, K'Q'}\,%
\hat{T}_{K''Q''}^{(a)}(t)%
\label{2.6}%
\end{eqnarray}%
For the interaction processes shown in figures \ref{Fig.1} and
\ref{Fig.2} only the following two combinations of the alignment
operators and the longitudinal orientation component contribute to
the dynamics of the atomic spin fluctuations
\begin{eqnarray}
\hat{T}_{xy}^{(a)}(t)&=&\frac{1}{2}\left[\hat{T}_{2-2}^{(a)}(t)\,+\,%
\hat{T}_{22}^{(a)}(t)\right]%
\nonumber\\%
\hat{T}_{\xi\eta}^{(a)}(t)&=&%
-\frac{1}{2i}\left[\hat{T}_{2-2}^{(a)}(t)\,-\,%
\hat{T}_{22}^{(a)}(t)\right]%
\nonumber\\%
\hat{F}_z^{(a)}(t)&=&\frac{1}{\sqrt{3}}\left[F_0(F_0+1)(2F_0+1)\right]^{1/2}\hat{T}_{10}^{(a)}(t)%
\label{2.7}%
\end{eqnarray}%
where $\hat{F}_{z}^{(a)}(t)$ is the Heisenberg operator of the
angular momentum projection for $a$-th atom on $z$-axis. For this
set of operators the commutation relation (\ref{2.6}) leads
\begin{eqnarray}
\left[\hat{T}_{xy}^{(a)}(t), \hat{T}_{\xi\eta}^{(b)}(t)\right]&=&%
\delta_{ab}\,ic_{1}\,\hat{F}_{z}^{(a)}(t)+\delta_{ab}\,ic_{3}\hat{T}_{30}^{(a)}(t)%
\nonumber\\%
\left[\hat{F}_{z}^{(a)}(t), \hat{T}_{xy}^{(b)}(t)\right]&=&%
\delta_{ab}\,2i\hat{T}_{\xi\eta}^{(a)}(t)%
\nonumber\\%
\left[\hat{F}_{z}^{(a)}(t), \hat{T}_{\xi\eta}^{(b)}(t)\right]&=&%
-\,\delta_{ab}\,2i\hat{T}_{xy}^{(a)}(t)%
\label{2.8}%
\end{eqnarray}%
The coefficients $c_{1}$ and $c_{3}$ are given by
\begin{eqnarray}
c_{1}&=&\frac{3}{F_0(F_0+1)(2F_0+1)}%
\nonumber\\%
c_{3}&=&-\frac{6\left[(F_0-1)(F_0+2)\right]^{1/2}}%
{\left[7F_0(F_0+1)(2F_0-1)(2F_0+1)(2F_0+3)\right]^{1/2}}%
\nonumber\\%
\label{2.8'}%
\end{eqnarray}%
and the higher-rank irreducible component $\hat{T}_{30}$ weighted
with coefficient $c_{3}$ contributes only for atoms with $F_0>1$.

The operators $\hat{T}_{xy}^{(a)}$ and $\hat{T}_{\xi\eta}^{(a)}$
have clear physical meaning and define the components of alignment
tensor with respect to either $x,y$ or $\xi,\eta$ Cartesian
frames, which were earlier introduced with definition of the
polarization Stokes components of light. These operators can also
be related with the transverse components of a quasi-spin, which
could be defined in the system of two Zeeman states coupled via
the $\Lambda$-type excitation channels shown in figures
\ref{Fig.1} and \ref{Fig.2}. However in context of our discussion
it is more important to emphasize the tensor nature of these
components, which are responsible for the fluctuations of linear
birefringence of the sample initiated by either stimulated or
spontaneous Raman processes shown in these figures.

The important step in further description of the collective
dynamics of the atomic spins consists in assumption that the
original discrete distribution of the point-like atomic spins can
be smoothed with procedure of the mesoscopic averaging. Then for a
mesoscopically thin layer located between $z$ and $z+\Delta z$ one
can define
\begin{eqnarray}
\hat{\mathcal T}_{xy}(z,t)&=&\frac{1}{\Delta z}%
\sum_{z<z_a<z+\Delta z}\,%
\hat{T}_{xy}^{(a)}(t)\,%
\nonumber\\%
\hat{\mathcal T}_{\xi\eta}(z,t)&=&\frac{1}{\Delta z}%
\sum_{z<z_a<z+\Delta z}\,%
\hat{T}_{\xi\eta}^{(a)}(t)\,%
\label{2.9}%
\end{eqnarray}%
where the sum is extended on the atoms located inside the layer.
These expressions define the smoothed spatial distribution of the
alignment components for which the first line of commutation
relation (\ref{2.8}) transforms to
\begin{eqnarray}
\lefteqn{\left[\hat{\mathcal T}_{xy}(z,t), \hat{\mathcal T}_{\xi\eta}(z',t)\right]=%
\delta(z-z')\,i\bar{c}_{13}\,\bar{\mathcal F}_{z}}%
\nonumber\\%
&&\bar{c}_{13}=\frac{15}{F_0(F_0+1)(2F_0+1)(2F_0+3)}%
\label{2.10}%
\end{eqnarray}%
where in the case of small fluctuation the right-hand side is
replaced by an expectation value for the density of atomic angular
momentum $\bar{\mathcal F}_{z}$. As in the previous section we
assume that this projection of the collective spin is
approximately conserved in the Raman process such that the total
angular momentum $F_{\Sigma}=\bar{\mathcal F}_{z}L=\pm F_0N_A$,
where $L$ is the sample length and $N_A$ is the number of atoms.
Since the incoherent losses are ignored $N_A$ remains the integral
of motion in this model.

\subsection{Dynamic equations}
The master equation governing the dynamics of atoms-field
variables in the Heisenberg picture can be derived similarly to
the way it was done in Ref. \cite{KMSJP}. We apply the effective
Hamiltonian derived in that paper and neglect the dissipation
channels caused by incoherent scattering.

We will assume that the squeezed probe radiation for the
excitation scheme, shown in figure \ref{Fig.1}, is characterized
by a degeneracy parameter (i.e. by a mean number of photons in the
coherence volume of the quantum radiation) higher than unity. As
was shown in Ref.\cite{KSS} there is a special scenario when the
$\Lambda$-system is probed by extremely weak SPDC light source
consisting of strongly correlated photon pairs, which should be
cooperatively scattered by all atoms of the ensemble. This process
cannot be described by the Hamiltonian of Ref.\cite{KMSJP} and
should be discussed separately.

Omitting the derivation details we present the following
Heisenberg wave-type equations describing the temporal and spatial
dynamics of the Stokes components and the alignment components
\begin{eqnarray}
\left[\frac{\partial}{\partial z}+\frac{1}{c}%
\frac{\partial}{\partial t}\right]%
\hat{\Xi}_1(z,t)&=&%
\kappa_1\,\hat{\Xi}_3(z,t)\;-\;%
2\epsilon\,\bar{\Xi}_2\,\hat{\mathcal T}_{xy}(z,t)%
\nonumber\\%
\left[\frac{\partial}{\partial z}+\frac{1}{c}%
\frac{\partial}{\partial t}\right]%
\hat{\Xi}_3(z,t)&=&%
-\kappa_1\,\hat{\Xi}_1(z,t)\;+\;%
2\epsilon\bar{\Xi}_2%
\hat{\mathcal T}_{\xi\eta}(z,t)\,%
\nonumber\\%
\frac{\partial}{\partial t}\hat{\mathcal T}_{xy}(z,t)&=&%
-\bar{\Omega}\hat{\mathcal T}_{\xi\eta}(z,t)\;+\;%
\bar{c}_{13}\epsilon\,\bar{\mathcal F}_{z}\,%
\hat{\Xi}_1(z,t)\,%
\nonumber\\%
\frac{\partial}{\partial t}\hat{\mathcal T}_{\xi\eta}(z,t)&=&%
\phantom{+}\bar{\Omega}\,\hat{\mathcal T}_{xy}(z,t)\;-\;%
\bar{c}_{13}\epsilon\,\bar{\mathcal F}_{z}\,%
\hat{\Xi}_3(z,t)\,%
\nonumber\\%
\label{2.11}%
\end{eqnarray}
These equations are written for the physical obsrevables and have
a rather clear physical structure. They show that in a linearized
regime the modification of the light and atomic polarization in a
forward passage results in a combine action of gyrotropy and
linear birefringence. In the discussed case the former manifests
itself as an averaged interaction between atoms and light and the
latter actually creates the quantum fluctuation interface between
these subsystems. Let us point out that in a formal description
based on a $\Lambda$-type configuration the fluctuations of
birefringence (alignment) components are usually treated as an
atomic coherence created by Raman process in a quasi-spin
subsystem and the importance of the phase-matching gyrotropy can
be underestimated. We show how this effect can be properly taken
into consideration by direct solution of Eqs.(\ref{2.11}).

Equations (\ref{2.11}) introduce the following important
parameters: $\kappa_1$, $\bar{\Omega}=2\Omega_0+\Omega_1$ and
$\epsilon$. The first two describe the changes in the mean
classical parameters of light and atoms. $\kappa_1$ describes the
gyrotropy of the sample caused by the average angular momentum
orientation of the atomic spins. $\Omega_1$ is the light shift
between the Zeeman sublevels $F_0,M=\pm F_0$ and $F_0,M=\pm F_0\mp
2$. It is additionally assumed that the atoms interact with an
external magnetic field and $\Omega_0$ is the relevant Zeeman
splitting for the neighboring sublevels. The third parameter
$\epsilon$ is the most important and is responsible for the
coupled dynamics of the fluctuations of the polarization
components of light and of the atomic alignment components. The
microscopic expressions for all these characteristics for the case
of hyperfine transitions of alkali atoms are summarized in
Appendix \ref{A}

Before we discuss the solutions of these equations we turn again
to the physics of the processes shown in figures \ref{Fig.1} and
\ref{Fig.2}. The difference between these processes is only
indicated by different signs of $\bar{\mathcal F}_{z}$ in the
second pair of equations (\ref{2.11}). For positive $\bar{\mathcal
F}_{z}$ (Fig.\ref{Fig.1}) these equations yield the normal
wave-type solution leading to the swapping of the quantum states
between the light and spin subsystems. Then the memory protocol
can develop in the following way. If the quantum field is prepared
in an unknown left-hand polarized squeezed-vacuum state its
momentum/position quadrature components can be expressed as
\begin{eqnarray}
X_{P}&\propto&-ie^{i\theta}a+ie^{-i\theta}a^{\dagger}%
\nonumber\\%
X_{Q}&\propto&e^{i\theta}a+e^{-i\theta}a^{\dagger}%
\label{2.12}%
\end{eqnarray}
where the phase $\theta$ determining the orientation of the
squeezed and anti-squeezed quadratures in the phase plane as well
as their variances are the unknown parameters of the state. If
such an unknown squeezed state is superimposed with a classical
coherent mode one straightforwardly gets that
\begin{eqnarray}
\hat{\Xi}_1&\propto &X_P%
\nonumber\\%
\hat{\Xi}_3&\propto &X_Q%
\label{2.13}%
\end{eqnarray}%
The crucial point for the atoms-field dynamics described by
equations (\ref{2.11}) and for the memory protocol is that the
direction of the axes $x,y$ and $\xi,\eta$ remains {\textit
completely unknown} in the experiment, see figure \ref{Fig.1}.
Thus the unknown squeezed state transforms into the unknown
polarization squeezing such that the $\hat{\Xi}_1$ component is
squeezed and $\hat{\Xi}_3$ is anti-squeezed. The dynamics of these
components is further developing in accordance with equations
(\ref{2.11}) and under certain conditions the squeezed state of
light can be converted into the squeezed state of the alignment
components of the atomic spin angular momenta.

For the negative sign of $\bar{\mathcal F}_{z}$ (Fig.\ref{Fig.2})
there is no normal wave solution of the equations (\ref{2.11}). As
we will show in this case there is an exponential enhancement of
the quantum fluctuations of the Stokes components and of the
atomic alignment. Such an anomalous spin polariton wave describes
an entangled state of these subsystems. After interaction the low
frequency modes of the polarization and intensity spectrum of the
outgoing light will be entangled with the alignment-type standing
wave modes of the atomic spins. Such process can be for example
utilized in the quantum repeater protocol with remote atomic
systems.

\section{Coupled dynamics of the atomic and field subsystems}

The Heisenberg equations (\ref{2.11}) can be solved via the method
of Laplace transform similar to how it was done in
Ref.\cite{KMSJP}. The solution and verification of the commutation
relations are given in Appendix \ref{B}. In this section we
discuss the physical consequences which are mainly important from
the quantum information point of view. For the sake of simplicity
and without loss of generality, see Ref.\cite{KMSJP}, we will
ignore the retardation effects and restrict our discussion to the
low frequency fluctuations such that in equations (\ref{2.11}) we
neglect time derivatives of the field variables. We also restrict
our consideration to the special case when the relevant pair of
Zeeman sublevels, which is either $|F_0,F_0\rangle$ and
$|F_0,F_0-2\rangle$ (Fig.\ref{Fig.1}) or $|F_0,-F_0\rangle$ and
$|F_0,-F_0+2\rangle$ (Fig.\ref{Fig.2}), is always degenerate such
that $\bar{\Omega}=2\Omega_0+\Omega_1=0$ during the interaction
and $\Omega_0=0$ without the interaction. This non-critical but
convenient simplification allows us to consider storage of quantum
states of light in the time independent standing spin wave without
any Zeeman-type oscillations, see appendix \ref{B}.

The qualitative character of the cooperative dynamics of the
atomic and field variables can be illustrated by the dispersion
relation (\ref{b.6}) for the spin polariton modes and light modes
defined in the semi-infinite medium and for infinite interaction
time. Then temporal and spatial dynamics of the process can be
expressed by the respective Laplace modes $s$ and $p$ introduced
by expansion (\ref{b.4}). The Laplace variables for the inverse
transform can be parameterized as $s=-i\Omega,p=iq$. Then the
inverse transform reveals a spectral Fourier expansion over the
set of relevant temporal and spatial modes. The wave-type dynamics
makes these modes coupled via the following dispersion law
\begin{eqnarray}
\Omega &=& \frac{A}{q}%
\nonumber\\%
A &=& -2\bar{c}_{13}\epsilon^2\bar{\Xi}_2\bar{\mathcal F}_z%
\label{3.1}%
\end{eqnarray}
This dispersion relation reflects the coupling of low frequency
temporal fluctuations of light to fine scale spatial fluctuations
of atoms and of fast temporal fluctuations of light to long scale
fluctuations of atoms respectively. For an atomic medium of the
length $L$ and a light pulse of the duration $T$, the combination
$ATL$, characterizing the overall coupling strength between light
pulse and atomic ensemble, quantifies this scale relation in
natural units.

\subsection{Quantum memory protocol}\label{IIIA}

In the following we apply the input/output relations derived in
appendix \ref{B} to identify the suitable parameters for writing
and retrieving to and from a quantum memory. For similar systems
the optimization of a quantum memory has been analyzed very
thoroughly in the single and few-mode situations
\cite{GAFSL,NWRSWWJ}, while here we are interested in a multi-mode
situation considering pulses of squeezed light of a duration much
longer than the inverse bandwidth of squeezing. Prior studies of
squeezed state storage in \cite{DBP} considered a steady state
regime, while here we discuss a temporal sequence of writing,
storage and retrieval.

\subsubsection{Write-in stage}
Consider the experimental scheme shown in figure \ref{Fig.1} for a
pulse-type excitation of the system with a portion of quantum
light with the following correlation properties
\begin{eqnarray}
\frac{1}{2}%
\langle\left\{\hat{\Xi}_1(\tau),\hat{\Xi}_1(0)\right\}_{+}\rangle%
&=&\left(\delta(\tau)+\xi_1\,\frac{1}{2\tau_1}\,%
e^{-\frac{|\tau|}{\tau_1}}\right)\bar{\Xi}_2%
\nonumber\\%
\frac{1}{2}%
\langle\left\{\hat{\Xi}_3(\tau),\hat{\Xi}_3(0)\right\}_{+}\rangle%
&=&\left(\delta(\tau)+\xi_3\,\frac{1}{2\tau_3}\,%
e^{-\frac{|\tau|}{\tau_3}}\right)\bar{\Xi}_2%
\nonumber\\%
\label{3.3}%
\end{eqnarray}%
where $\{\ldots,\ldots\}_{+}$ denotes the anti-commutator of two
observables. These correlation functions describe the output
generated by the intra-cavity subthreshold degenerate parametric
amplifier, see Ref.\cite{GrdCll}. The freely propagating light
with such correlation properties can be associated with a squeezed
state of a harmonic oscillator where $\Xi_1$ component is squeezed
($\xi_1<0$) and $\Xi_3$ component is anti-squeezed ($\xi_3>0$)
such that $(1+\xi_1)(1+\xi_3)=1$. The spectral bandwidth of the
informative part of the quantum light is limited by the longest of
two correlation times
\begin{eqnarray}
\tau_1&=&\left[\frac{\gamma_C}{2}+\kappa_D\right]^{-1}%
\nonumber\\%
\tau_3&=&\left[\frac{\gamma_C}{2}-\kappa_D\right]^{-1}%
\label{3.4}%
\end{eqnarray}
where $\gamma_C$ is the cavity loss rate through the output mirror
and $\kappa_D$ is the efficiency of the downconversion process.
For high level of squeezing one has $\tau_3\gg\tau_1$. This
inequality indicates that the minimal duration $T$ of the quantum
light pulse should be considerably longer than the longest time
$\tau_3\equiv\tau_c$.

The general solution (\ref{b.8}), given in the appendix \ref{B},
can be straightforwardly rewritten in the following form
\begin{eqnarray}
\lefteqn{\hat{{\mathcal T}}_{\mathrm{I}}^{\mathrm{out}}(z)\equiv%
\cos\kappa_1z\,\hat{{\mathcal T}}_{xy}(z,T)+\sin\kappa_1z\,\hat{{\mathcal T}}_{\xi\eta}(z,T)=}%
\nonumber\\%
&&c_{13}\epsilon\bar{\cal F}_z\int_0^T\!dt\,%
J_0\!\left(2[-A(T-t)z]^{1/2}\right)\,\hat{\Xi}_1^{\mathrm{in}}(t)+\ldots%
\nonumber\\%
\nonumber\\%
\lefteqn{\hat{{\mathcal T}}_{\mathrm{I\!I\!I}}^{\mathrm{out}}(z)\equiv%
\cos\kappa_1z\,\hat{{\mathcal T}}_{\xi\eta}(z,T)-\sin\kappa_1z\,\hat{{\mathcal T}}_{xy}(z,T)=}%
\nonumber\\%
&&-c_{13}\epsilon\bar{\cal F}_z\int_0^T\!dt\,%
J_0\!\left(2[-A(T-t)z]^{1/2}\right)\,\hat{\Xi}_3^{\mathrm{in}}(t)+\ldots%
\nonumber\\%
\label{3.5}%
\end{eqnarray}
which shows how the input field operators are transferred into
space-dependent atomic spin operators. The dots in the right hand
side stand for the contributions of the atomic operators
responsible for reproduction of input atomic state, which make
this transformation imperfect.

The principle condition, which makes the memory protocol feasible,
is that for the low-frequency spatial fluctuations on the left
hand side of Eqs.(\ref{3.5}) the contribution of the input spin
fluctuations on the right hand side is suppressed. In the spectral
expansion of the operators $\hat{{\mathcal
T}}_{\mathrm{I}}^{\mathrm{out}}(z)$ and $\hat{{\mathcal
T}}_{\mathrm{I\!I\!I}}^{\mathrm{out}}(z)$ such fluctuations belong
to the spectral domain $q\lesssim q_c=[\frac{|A|T}{L}]^{1/2}$. In
accordance with the dispersion law (\ref{3.1}), which is
asymptotically valid for $T,L\to \infty$, the transform
(\ref{3.5}) performs mapping of the field fluctuations within the
spectral interval $0<\Omega\lesssim\tau_c^{-1}$ onto the spatially
dependent spin fluctuations within the spatial spectral domain
$\infty>q\gtrsim |A|\tau_c$. For reliable mapping of the field
quantum state onto the atomic alignment subsystem the latter
domain should essentially overlap with the spectral domain where
the input spin fluctuations are suppressed. For the efficient
memory protocol the following inequalities should be fulfilled
\begin{eqnarray}
|A|\tau_c&\ll & q_c=\left[\frac{|A|T}{L}\right]^{1/2}%
\nonumber\\%
L^{-1}&\ll & q_c%
\label{3.6}%
\end{eqnarray}
These inequalities imply that the contribution of the input spin
fluctuations are suppressed and ensure that the spatial
fluctuations of $\hat{{\mathcal
T}}_{\mathrm{I}}^{\mathrm{out}}(z)$ and $\hat{{\mathcal
T}}_{\mathrm{I\!I\!I}}^{\mathrm{out}}(z)$ correctly reproduce  the
temporal dynamics of the input field fluctuations (\ref{3.3}) in
the relevant part of their spectrum. However the dispersion
relation (\ref{3.1}) has a singular behavior for the function
$\Omega=\Omega(q)$ near the point $q\to 0$ such that the shot
noise part of the input fluctuation spectrum is reproduced at the
zero point of spatial spectrum. Thus near the spectral point $q\to
0$, associated with an integral collective mode of the standing
spin waves in the limit $L\to\infty$, the informative part of the
fluctuation spectrum will be invisible. However, for the sample
with a finite length $L$ the squeezed state can be mapped onto the
integral collective modes of $\hat{{\mathcal
T}}_{\mathrm{I}}^{\mathrm{out}}(z)$ and $\hat{{\mathcal
T}}_{\mathrm{I\!I\!I}}^{\mathrm{out}}(z)$ if $|A|\tau_c\ll
L^{-1}\ll q_c$ i.e. for a spectrally broad incoming quantum light.

\subsubsection{Retrieval stage}

For retrieval of the quantum state back onto light the atomic
ensemble should be probed with another strong coherent light
pulse. Optimal retrieval occurs when
\begin{equation}
\kappa_1L=2\pi\times\mathrm{any\ integer}%
\label{3.7}%
\end{equation}
Then following (\ref{b.7}) the output field operators are given by
\begin{eqnarray}
\hat{\Xi}_{1}^{\mathrm{out}}(t)\!\!&=&\!\!%
-2\epsilon\bar{\Xi}_2'\!\int_0^L\!dz\,%
J_0\!\left(2[-A'(L-z)t]^{1/2}\right)\hat{{\mathcal T}}_{\mathrm{I}}^{\mathrm{out}}(z)+\ldots%
\nonumber\\%
\hat{\Xi}_{3}^{\mathrm{out}}(t)\!\!&=&\!\!%
\phantom{+}2\epsilon\bar{\Xi}_2'\!\int_0^L\!dz\,%
J_0\!\left(2[-A'(L-z)t]^{1/2}\right)\hat{{\mathcal T}}_{\mathrm{I\!I\!I}}^{\mathrm{out}}(z)+\ldots%
\nonumber\\%
\label{3.8}%
\end{eqnarray}
where the initial state of the atoms is $\hat{{\mathcal
T}}_{\mathrm{I}}^{\mathrm{out}}(z)$ and $\hat{{\mathcal
T}}_{\mathrm{I\!I\!I}}^{\mathrm{out}}(z)$ modified in the write-in
stage of the protocol. The dots on the right hand side again
indicate the imperfection of the transform because of the presence
of the input field operators associated with the vacuum modes.

The noise contribution caused by the vacuum fluctuation of the
input field can be suppressed in the low frequency domain
$\Omega<\Omega_c'=[\frac{|A'|L}{T'}]^{1/2}$, see Eq.(\ref{b.7}).
At the retrieval stage of the protocol the spin fluctuations
within the spectral interval $0<q\lesssim q_c$ are mapped onto the
time dependent field fluctuations in the frequency spectral domain
$\infty>\Omega\gtrsim |A'|l_c$, where the correlation length is
given by $l_c=q_c^{-1}\ll L$. At the retrieval stage of the memory
protocol the following inequalities should be fulfilled
\begin{eqnarray}
|A'|l_c&\ll & \Omega_c'=\left[\frac{|A'|L}{T'}\right]^{1/2}%
\nonumber\\%
T'^{-1}&\ll & \Omega_c'%
\label{3.9}%
\end{eqnarray}
which, after the obvious replacement of temporal and spatial
parameters, are symmetric to the inequalities (\ref{3.6}) and have
similar physical meaning.  Thus the time dynamics of the output
field operators $\hat{\Xi}_{1}^{\mathrm{out}}(t)$ and
$\hat{\Xi}_{3}^{\mathrm{out}}(t)$ correctly reproduce the spatial
distribution of spin fluctuations in the relevant part of the
spectrum.

\subsubsection{Physical requirements and numerical simulations}

Basic requirements for the proposed memory protocol can be
summarized by combining the inequalities (\ref{3.6}) and
(\ref{3.9}) where parameters $A$ and $A'$ are given by
Eq.(\ref{3.1}). The following limitation should be imposed on the
number of atoms $N_A$ and on the numbers of photons in the
coherent strong field $N_P$ and $N_P'$ participating in the
process  at the write-in and retrieval stages respectively
\begin{eqnarray}
\epsilon^2N_AN_P&\gg& 1%
\nonumber\\%
\epsilon^2N_AN_P'&\gg& 1%
\label{3.10}%
\end{eqnarray}
The substitution of $l_c$ from (\ref{3.6}) into (\ref{3.9}) yields
\begin{equation}
N_P'\ll N_P%
\label{3.11}%
\end{equation}
From evaluation of incoherent losses we obtain the following
conditions on the numbers of atoms and photons
\begin{eqnarray}
N_A\sigma_{F_{0}}^{-}&\ll& S_0%
\nonumber\\%
N_P\sigma_{F_{0}-2}^{+},\,N_P'\sigma_{F_{0}-2}^{+}&\ll& S_0%
\label{3.12}%
\end{eqnarray}
where $\sigma_{F_{0}}^{-}$ is the cross section of the incoherent
scattering for a left-hand polarized light from the atoms
populating the Zeeman sublevel $M=F_0$ and $\sigma_{F_{0}-2}^{+}$
is the cross section of the incoherent scattering for a right-hand
polarized light from the atoms populating the $M=F_0-2$ Zeeman
sublevel; $S_0$ is the beam/sample cross area. The first line
indicates that the atomic medium should be transparent for the
informative quantum radiation. The second line indicates that the
incoherent scattering should have negligible influence on the
dynamics of the spin coherence during both the write-in and the
retrieval cycles. These requirements lead to the following basic
demand on the density of atoms $n_0$, resonance radiation
wavelength $\lambdabar$ and the sample length $L$:
$n_0\lambdabar^2L\gg 1$, i.e. the on resonance optical depth of
the sample needs to be large

Optimization of the inequalities (\ref{3.11}) and (\ref{3.12})
requires that $N_P'\leq N_A\leq N_P$. To explain this choice let
us again consider the coherent Raman scattering as the physical
background for the memory protocol. In the write-in stage of the
protocol the most desirable process is the annihilation of the
photons of the quantum modes as a result of the coherent
scattering into the classical mode followed by the transfer of
their quantum state onto the alignment of atoms. The swapping
process can be easier done for the photons in the wings of the
spectral distribution (\ref{3.3}) and is more difficult to realize
for the resonance photons. That is why the process runs more
efficiently with a stronger coherent field with $N_P\geq N_A$. At
the retrieval stage the coherent Raman scattering takes photons
from the strong mode into the quantum modes. The spin fluctuations
of the atomic alignment tensor transfer back into the outgoing
modes of the quantum light. To make this process more efficient,
i.e. to increase the number of scattering events, it is desirable
that the sample is extended and contains a large number of the
atoms such that $N_A\geq N_P'$.

\begin{figure}[tp]
\vspace{\baselineskip}
\includegraphics{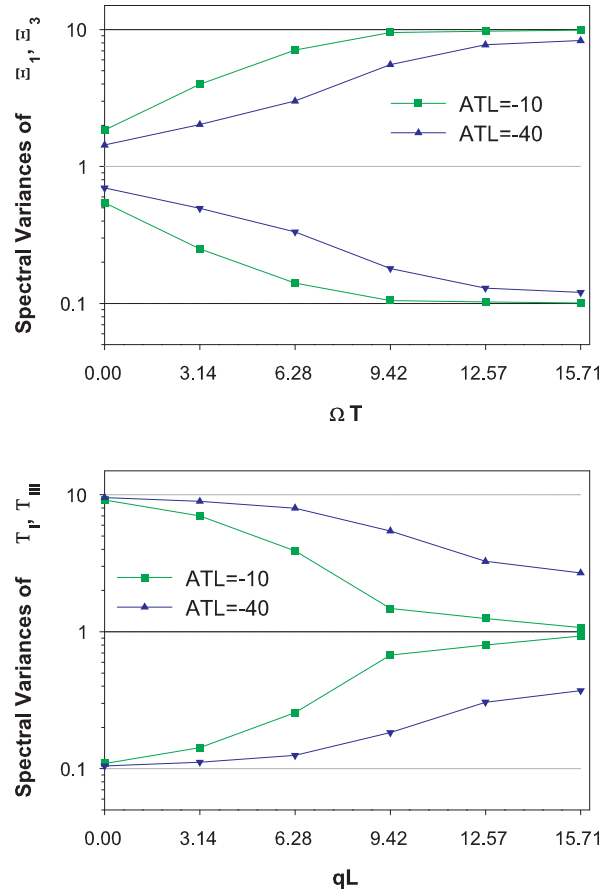}
\caption{The spectral variances of the Stokes components (upper
panel) and of the atomic alignment components (lower panel) before
and after the interaction with the broadband squeezed light for
cooperative parameter $ATL=-10$ (squares) and $ATL=-40$
(triangles). The black solid lines indicate the original
fluctuation spectra in the light and spin subsystems. The gray
lines in each of the panels indicate the original spectra for a
complementary system.}
\label{Fig.3}%
\end{figure}

In figure \ref{Fig.3} we show how the spectral variances of the
polarization components of light and atoms are modified after
sending the broadband squeezed light ($\tau_c\to 0$) through the
atomic sample. The graphs clearly display the swapping mechanism
at the write-in stage of the memory protocol. The graphs are
normalized to the vacuum state variance and the deviations from it
are expressed by the respective Mandel parameters $1+\xi_{\mathrm
i}$ for ${\mathrm i}=1,3$, which depend on either frequency
$\Omega$ (for light) or wave number $q$ (for atoms). The input
squeezed state is described by two spectrally independent
parameters $1+\xi_3^{\mathrm{in}}=10$ and
$1+\xi_1^{\mathrm{in}}=0.1$. The suppression of the correlations
in the low frequency domain of the temporal spectrum of light is
compensated by the enhancement of the correlations in the atoms
for the low frequency part of the spatial spectrum. The dominant
role of the collective modes in this process is a result of the
approximation $\tau_c\to 0$.

The data, presented in these graphs, corresponds to the red
detuning of light near the $D_1$-line of ${}^{87}$Rb for atoms in
the hyperfine level $F_0=1$ and for two selected values of the
cooperative interaction parameter $ATL=-10,-40$. The first value
corresponds to the detuning $-205\,$MHz from $F_0=1\to F=1$
resonance, where $\kappa_1=0$ (no gyrotropy effect) and for
approximately ten percent losses estimated from Eqs.(\ref{3.12}).
The second number is achievable for the detuning of a few thousand
MHz in the red wing of the $F_0=1\to F=1$ transition and for the
same level of losses.

\begin{figure}[tp]
\vspace{\baselineskip}
\includegraphics{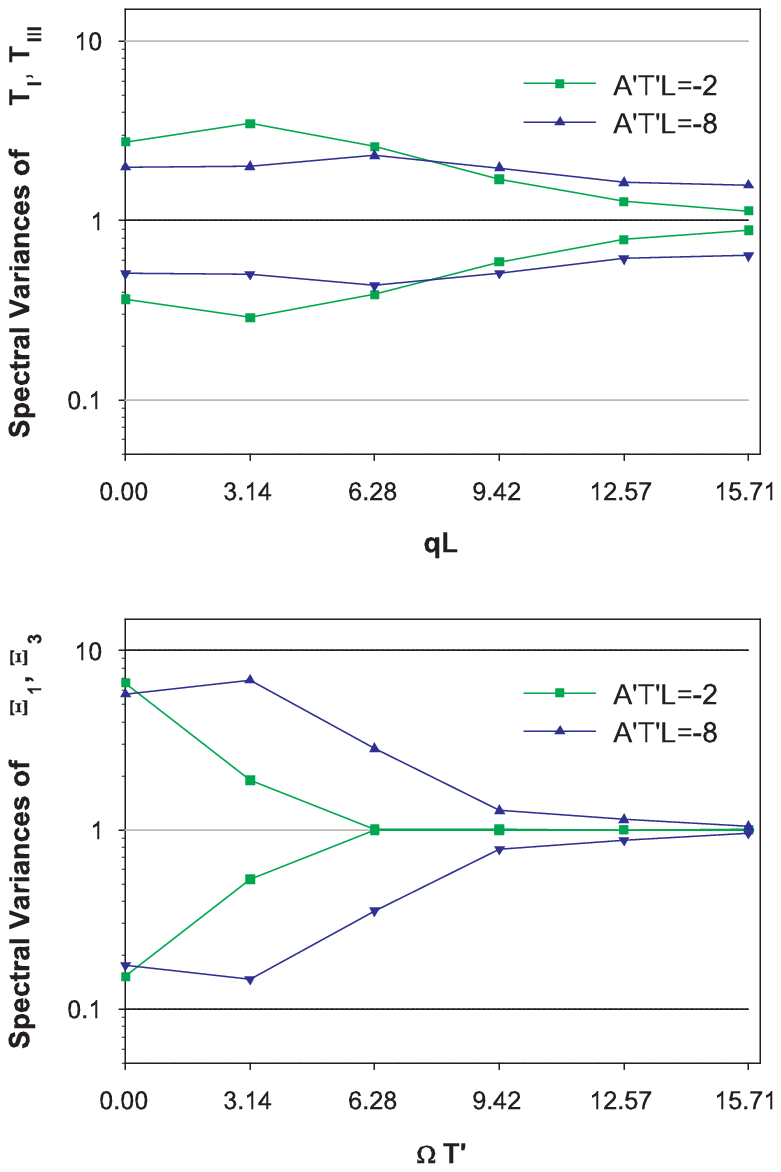}
\caption{The spectral variances of the alignment components of the
atoms (upper panel) and of the retrieved Stokes components of the
light (lower panel) after the readout of the state stored in the
atoms at the write-in stage of the memory protocol, see
Fig.\ref{Fig.3}. The values of cooperative parameter $A'T'L=-2$
(squares) and $A'T'L=-8$ (triangles) are coordinated with the data
of Fig.\ref{Fig.3}. The black solid lines indicate the original
fluctuation spectra in the spin and light subsystems. The gray
lines in each of the panels indicate the original spectra for the
complementary system.}
\label{Fig.4}%
\end{figure}

Figure \ref{Fig.4} shows the spectral variances of the
polarization components for the atoms and light at the retrieval
stage of the protocol. The figure shows how well the recovered
state can reproduce the input. The parameters are chosen such that
$A'T'L=-2$ corresponding to $ATL=-10$ and $A'T'L=-8$ corresponding
to $ATL=-40$. Let us point out once more that for the best overall
efficiency, the number of photons in the strong coherent pulse at
the retrieval stage should be smaller than the number of photons
applied at the write-in stage of the protocol and smaller than the
number of atoms. As follows from these results the retrieved
quantum state of light can reproduce the input state only in
certain parts of the fluctuation spectrum. Further optimization
should allow for identification of the best temporal mode, where
the retrieval of the original squeezed state would be optimal.

\begin{figure}[tp]
\vspace{\baselineskip}
\includegraphics{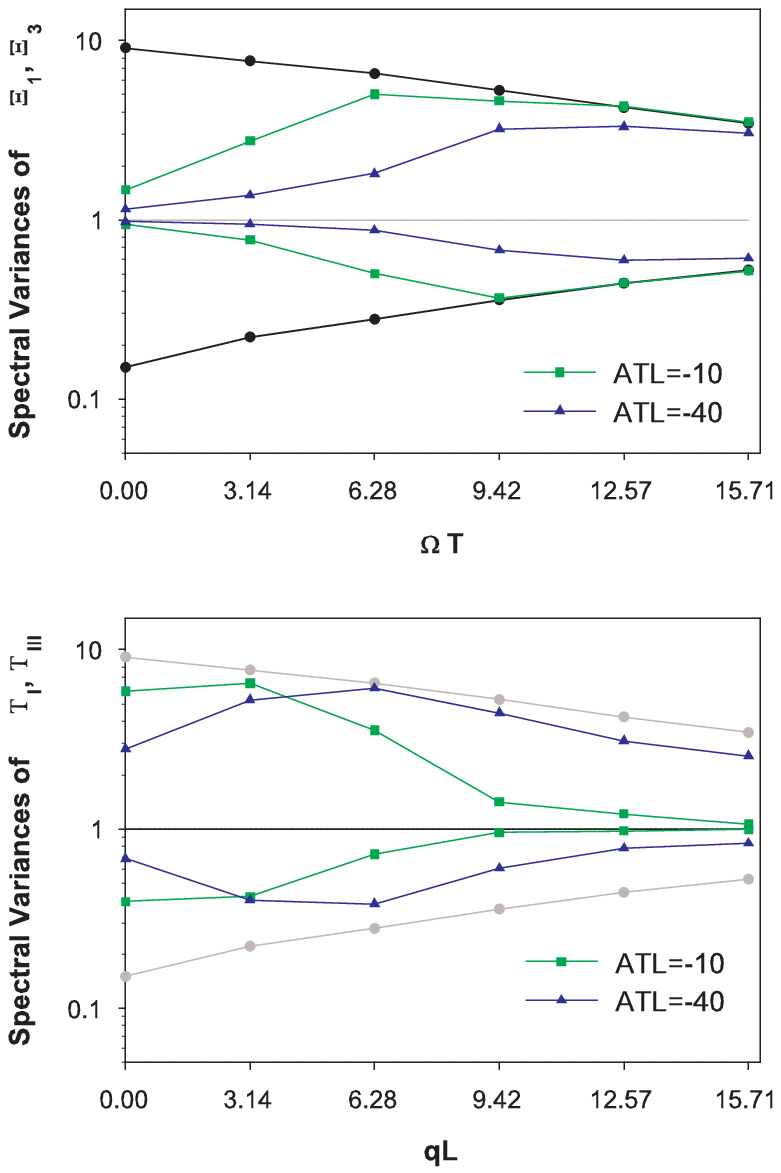}
\caption{Same as in Fig.\ref{Fig.3}, but for the finite-bandwidth
squeezed light. The ratio of interaction time to the correlation
time of squeezing for these graphs is $T/\tau_c=10$.}
\label{Fig.5}%
\end{figure}

\begin{figure}[tp]
\vspace{\baselineskip}
\includegraphics{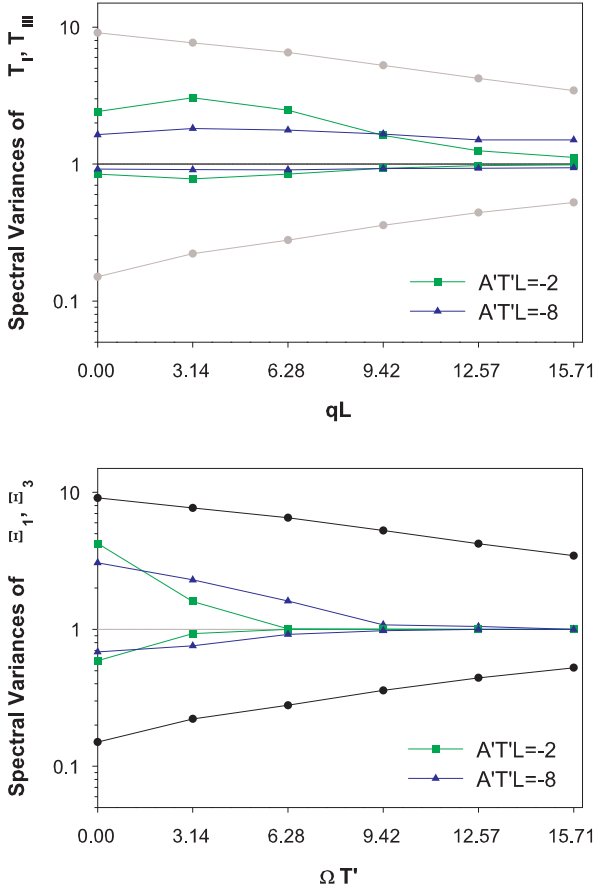}
\caption{Same as in Fig.\ref{Fig.4}, but for retrieval of the
finite-bandwidth squeezed state mapped onto the spin subsystem,
see Fig.\ref{Fig.5}. The original spectrum for the squeezed light
in the lower panel relates to the case of $T'=T$.}
\label{Fig.6}%
\end{figure}

Figures \ref{Fig.5} and \ref{Fig.6} show how the spectra of the
light and atoms are modified at the write-in and retrieval stages
with the finite-bandwidth squeezed light as the input. The
dependencies plotted in these graphs illustrate the importance of
the dispersion relation between the temporal and spatial modes
participating in the protocol. The ratio of the interaction time
to the correlation time of the squeezing is taken to be
$T/\tau_c=10$. As follows from the displayed results, for the
finite-bandwidth squeezed light the integral collective modes are
not optimal in both storage and retrieval steps of the protocol. A
certain optimization procedure is necessary for the best encoding
of the quantum information into particular domains and modes of
the fluctuation spectra for the light and atoms. Another important
observation which follows from these graphs is that the finite
bandwidth squeezing is more difficult to store and retrieve than
the broadband input state, the fact already established for the
storage phase in \cite{KMP}. This is the direct consequence of the
imperfection of the swapping mechanism for mapping the low
frequency fluctuations, as follows from the dispersion law
discussed above.

\subsubsection{Fidelity}

Fidelity is the fundamental criterion for any quantum memory or
teleportation scenario. It can be relatively simply defined for a
single mode situation and for a pure input state, and is more
subtle for a general case. In the following we show qualitatively
that in the present case the quantum scheme has always a better
fidelity than for what can be called a classical
memory/measurement protocol.

\begin{figure}[tp]
\vspace{\baselineskip}
\includegraphics{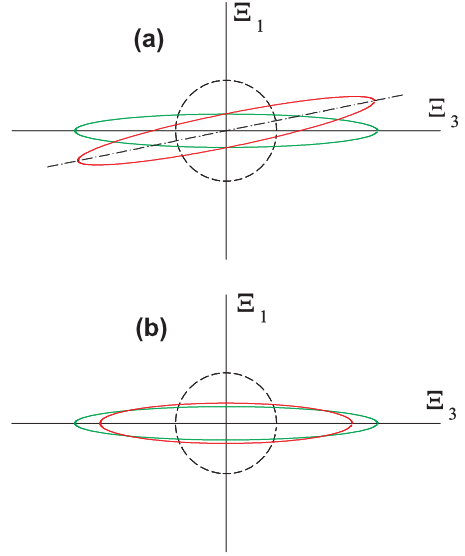}
\caption{The overlap of the Wigner function for the input squeezed
state (green ellipse) with its "classical" (a) and "quantum" (b)
counterparts (red ellipses).}
\label{Fig.7}%
\end{figure}

We assume that the classical memory scheme utilizes a balance
homodyne detection of the input squeezed state for the measurement
of its parameters. The measurement should yield the degree of
squeezing and the directions of the linear polarizations shown in
figure \ref{Fig.1}. Such a procedure is displayed in figure
\ref{Fig.7}a as an overlap of the Wigner function associated with
the input state with its measured counterpart. This overlap in the
limit of high level of squeezing can be expressed by the following
fidelity
\begin{equation}
F\approx \frac{1}{\sqrt{1+(D_3\theta_N})^2}%
\label{3.12'}
\end{equation}
where $D_3=(1+\xi_3)/2$ is half of the variance for the
anti-squeezed $\Xi_3$ component, which is assumed to be reliably
defined, and $\theta_N\sim\pi/N$ is the angular uncertainty
remaining after $N$ measuring attempts to identify direction of
$x,y$ or $\xi,\eta$ axes. Following Ref.\cite{KuprSk} the
following constraint should be fulfilled in order to make the
classical protocol efficient
\begin{equation}
\sqrt{\frac{\tau_c}{T_N}}<(D_3\theta_N)^2<1%
\label{3.12''}
\end{equation}
where $T_N\sim T/N$ is the duration of each measuring attempt as a
fraction of the total measuring time $T$. It is clear that for
$D_3\to\infty$ the number of measurements $N$ should grow up to
infinity such that for any limited time $T$ the constraint
(\ref{3.12''}) can never be fulfilled.

On the contrary, in the quantum memory protocol it is only
necessary to effectively reproduce the variances associated with
the squeezed and anti-squeezed polarization components. The
directions of the $x,y$ and $\xi,\eta$ axes remain completely
unknown in the interaction process. This is shown in figure
\ref{Fig.7}b with a more effective overlap than in case of figure
\ref{Fig.7}a. The fidelity can be conveniently written in terms of
Mandel parameter for the input and output states
\begin{equation}
F=\frac{2}{\left[(2+\xi_1^{\mathrm{in}}+\xi_1^{\mathrm{out}})%
(2+\xi_3^{\mathrm{in}}+\xi_3^{\mathrm{out}})\right]^{1/2}}%
\label{2.12'''}
\end{equation}
where "out" characteristics should be associated with optimally
defined spatial mode. Comparing (\ref{3.12'}) and (\ref{2.12'''})
for the same ratio $T/\tau_c$ and for the high level of squeezing
one can expect that in the optimized case the fidelity for the
quantum protocol will be always higher than the relevant classical
benchmark.

\subsection{Atoms-field entanglement}

The entanglement between the light and the atomic alignment
subsystems created in the process shown in figure \ref{Fig.2} is
attained in a result of the transformations (\ref{b.12}),
(\ref{b.13}) under the constraint (\ref{3.7}). The entanglement
can be written in the following canonic form
\begin{eqnarray}
\int_0^T h(t)\,\hat{\Xi}_1^{\mathrm{out}}(t)\,dt - %
\int_0^L g(z)\,\hat{{\mathcal T}}_{\mathrm{I}}^{\mathrm{out}}(z)\,dz &\to& 0%
\nonumber\\%
\int_0^T h(t)\,\hat{\Xi}_3^{\mathrm{out}}(t)\,dt + %
\int_0^L g(z)\,\hat{{\mathcal T}}_{\mathrm{I\!I\!I}}^{\mathrm{out}}(z)\,dz &\to& 0\phantom{3.13}%
\label{3.13}%
\end{eqnarray}%
The alignment spin waves $\hat{{\mathcal
T}}_{\mathrm{I}}^{\mathrm{out}}(z)$ and $\hat{{\mathcal
T}}_{\mathrm{I\!I\!I}}^{\mathrm{out}}(z)$ are defined by the upper
lines of Eqs.(\ref{3.5}). The standard entanglement between the
quadrature components of certain collective field and atomic spin
modes can be recognized in Eqs.(\ref{3.13}) , see also relations
(\ref{2.12}) and (\ref{2.13}).

Formally the temporal mode $h(t)$ and the spatial mode $g(z)$ can
be found by solving the following integral equations
\begin{eqnarray}
h(t)+ \int_t^T dt'\,\left[\frac{AL}{t'-t}\right]^{1/2}%
I_1\left(2\left[AL(t'-t)\right]^{1/2}\right)\,h(t')&&%
\nonumber\\%
- \bar{c}_{13}\epsilon\bar{\mathcal{F}}_z%
\int_0^L dz\,I_0\left(2\left[A(T-t)z\right]^{1/2}\right)\,g(z)\to 0\phantom{(3...14)}&&%
\nonumber\\%
\nonumber\\%
g(z)+ \int_z^L dz'\,\left[\frac{AT}{z'-z}\right]^{1/2}%
I_1\left(2\left[AT(z'-z)\right]^{1/2}\right)\,g(z')&&%
\nonumber\\%
+ 2\epsilon\bar{\Xi}_2%
\int_0^T dt\,I_0\left(2\left[A(L-z)t\right]^{1/2}\right)\,h(t)\to 0\phantom{(3...14)}&&%
\label{3.14}%
\end{eqnarray}%
Due to the presence of non-symmetric integral operators in the
upper lines of these equations their actual solution does not
responsibly exist in terms of real functions. One can search for a
pair of real functions $h(t)$ and $g(z)$ which minimize the
absolute value of the left hand side of (\ref{3.14}). This
circumstance is reflected by the "limit to zero" in the right hand
side, which can be approximately approached for certain optimal
observation conditions.

It is intuitively clear that this can be achieved for a system
consisting of a large number of atoms
($\epsilon|\bar{\mathcal{F}}_z|L\gg 1$) and of a large number of
coherent photons ($\epsilon\bar{\Xi}_2T\gg 1$). Using the common
properties of the Fredholm-type equations the solutions for the
spatial and temporal modes of (\ref{3.13}) can be found in terms
of the eigenfunctions of the integral equations (\ref{3.14}).

\section{Conclusion}
We discuss the polarization sensitive interaction between the
Stokes components of light and the alignment components of the
atomic ensemble as a resource for quantum information interface.
Our model accurately describes the quantum nature of interaction
between the spin subsystem of ultracold alkali atoms and a plain
light wave propagating through the sample, in the absence of
losses. The method has two important advantages. First, the
process does not require any special manipulations like a feedback
or adjustment of the driving classical light. This creates good
outlooks for its experimental implementation. Second, the model
yields an analytical solution under rather general assumptions.
This allows for a convenient and clear interpretation of the
Heisenberg dynamics of atoms and field at every step of the
discussed quantum information protocols.

We consider two basic protocols: quantum memory and quantum
entanglement. The memory protocol is an example of the swapping
mechanism between the atomic and field subsystems. The quantum
information, which is originally encoded in the polarization
degrees of freedom of the light wave, can be mapped onto the spin
standing wave associated with atomic alignment components. However
because of the multimode nature of the interaction process the
quantum information can be spread among a number of spatial
spectral modes. We show how the protocol, particularly at the
final retrieval stage, can be optimized. As an important step of
such optimization we have comprehensively discussed the
imperfection of the quantum memory channel and made qualitative
comparison of its figure of merit with a competing classical
memory/measurement scheme.

The second protocol is generation of entanglement in the spin
oriented atomic ensemble probed with a circularly polarized
coherent light mode. Forward scattered Raman photons appear
strongly correlated with an alignment-type coherence of atomic
scatterers. When the process becomes extended in space as well as
in time it creates pairs of entangled temporal and spatial modes.
We introduced the system of two integral equations, whose solution
defines the structure of these modes.

\section*{Acknowledgments}
We would like to thank Dr. Igor Sokolov for fruitful discussion.
The work was supported by the Russian Foundation for Basic
Research (RFBR-05-02-16172-a), by INTAS (project ID: 7904) and by
the European grants within the networks COVAQIAL and QAP. O.S.M.
would like to acknowledge the financial support from the charity
Foundation "Dynasty". D.V.K. would like to acknowledge financial
support from the Delzell Foundation, Inc.

\appendix
\section{Coupling parameters of the Heisenberg equations}\label{A}

The wave-type Heisenberg equations (\ref{2.11}) straightforwardly
follow from the commutation relations between the operators of the
respective quantum observables with an effective Hamiltonian
responsible for the process of coherent forward scattering. The
latter was derived in Ref.\cite{KMSJP}. The exact equations are
linearized with assumption of small fluctuations when
$\bar{\Xi}_2$ and $\bar{\mathcal F}_z$ are approximately conserved
and interpreted as classical values. Then the gyrotropy constant
is given by
\begin{equation}
\kappa_1=\sum_{F}\bar{\alpha}_{F_0F}^{(1)}(\bar{\omega})%
\frac{\sqrt{3}\,\bar{\mathcal F}_z}%
{\left[F_0(F_0+1)(2F_0+1)\right]^{1/2}}%
\label{a.1}
\end{equation}
Here $\bar{\alpha}_{F_0F}^{(1)}(\bar{\omega})$ is the
dimensionless orientation income into the polarizability tensor
for the $F_0\to F$ hyperfine transition
\begin{eqnarray}
\bar{\alpha}_{F_0F}^{(1)}(\bar{\omega})&=&%
(-1)^{F+F_0}\frac{1}{\sqrt{2}}\,%
\left\{\begin{array}{ccc}1&1&1\\F_0&F_0&F\end{array}\right\}\,%
\nonumber\\%
&&\times\frac{4\pi\bar{\omega}}{S_0\,c}\,%
\frac{|d_{F_0F}|^2}{-\hbar(\bar{\omega}-\omega_{FF_0})}%
\label{a.2}%
\end{eqnarray}%
where $d_{F_0F}$ is the reduced dipole moment and $\omega_{FF_0}$
is the transition frequency. The light shift is given by
\begin{equation}
\Omega_1=\sum_{F}\bar{\alpha}_{F_0F}^{(1)}(\bar{\omega})%
\frac{\sqrt{3}\,\bar{\Xi}_2}%
{\left[F_0(F_0+1)(2F_0+1)\right]^{1/2}}%
\label{a.3}
\end{equation}
This expression is similar to definition of the gyrotropy constant
(\ref{a.1}) because both the parameters come from the same
Faraday-type interaction term of the effective Hamiltonian.

The coupling constant $\epsilon$ responsible for the
alignment-type interaction is given by
\begin{equation}
\epsilon=\frac{1}{2}\sum_{F}\bar{\alpha}_{F_0F}^{(2)}(\bar{\omega})%
\label{a.4}
\end{equation}
where $\bar{\alpha}_{F_0F}^{(2)}(\bar{\omega})$ is the
dimensionless alignment contribution into the polarizability
tensor
\begin{eqnarray}
\bar{\alpha}_{F_0F}^{(2)}(\bar{\omega})&=&%
(-1)^{1+F+F_0}%
\left\{\begin{array}{ccc}1&1&2\\F_0&F_0&F\end{array}\right\}\,%
\nonumber\\%
&&\times\frac{4\pi\bar{\omega}}{S_0\,c}\,%
\frac{|d_{F_0F}|^2}{-\hbar(\bar{\omega}-\omega_{FF_0})}%
\label{a.5}%
\end{eqnarray}%
All the parameters are the spectrally dependent values such that
$\kappa_1=\kappa_1({\bar{\omega}})$,
$\Omega_1=\Omega_1({\bar{\omega}})$,
$\epsilon=\epsilon({\bar{\omega}})$ and this dependence can be
very important in practical calculations.

\section{The solution of the Heisenberg equations}\label{B}

We derive the solution for those physical conditions when only low
frequency temporal fluctuations are considered as a quantum
information carrier. That lets us completely ignore the
retardation effects associated with a finite sample size.
Practically this means that we can neglect the time derivations in
the first two lines of the system (\ref{2.11}).

As a first step we make the following local rotational
transformations for the Stokes components of the field subsystem
\begin{eqnarray}
\hat{\Xi}_{\mathrm I}(z,t)&=&\cos\varphi(z,t)\,\hat{\Xi}_{1}(z,t)-%
\sin\varphi(z,t)\,\hat{\Xi}_{3}(z,t)%
\phantom{(B1)}\nonumber\\%
\hat{\Xi}_{\mathrm{I\!I\!I}}(z,t)&=&\sin\varphi(z,t)\,\hat{\Xi}_{1}(z,t)+%
\cos\varphi(z,t)\,\hat{\Xi}_{3}(z,t)%
\label{b.1}
\end{eqnarray}
and for the alignment components of the atomic subsystem
\begin{eqnarray}
\hat{{\mathcal T}}_{\mathrm{I}}(z,t)\!\!&=&\!\!\cos\varphi(z,t)\hat{{\mathcal T}}_{xy}(z,t)+%
\sin\varphi(z,t)\hat{{\mathcal T}}_{\xi\eta}(z,t)%
\phantom{\ (B.2)}\nonumber\\%
\hat{{\mathcal T}}_{\mathrm{I\!I\!I}}(z,t)\!\!&=&\!\!-\sin\varphi(z,t)\hat{{\mathcal T}}_{xy}(z,t)+%
\cos\varphi(z,t)\hat{{\mathcal T}}_{\xi\eta}(z,t)%
\label{b.2}
\end{eqnarray}
where $\varphi(z,t)=\kappa_1z\!+\!\bar{\Omega}t$. Without losing
of generality the parameter $\bar{\Omega}$ can be set as $0$ to
ignore any non-principle freely precession of atomic spins
associated with external magnetic field and light shift. It is
also convenient to set the sample length $L$ as
$\kappa_1L=(2\pi\times \mathrm{any\ integer})$, then the pair of
Stokes variables $\hat{\Xi}_{\mathrm I}(z,t)$,
$\hat{\Xi}_{\mathrm{I\!I\!I}}(z,t)$ and $\hat{\Xi}_{1}(z,t)$,
$\hat{\Xi}_{3}(z,t)$ will coincide at the output of the sample.

Applying the orthogonal transformations (\ref{b.1}) and
(\ref{b.2}) to the system (\ref{2.11}) the latter can be rewritten
in the following form
\begin{eqnarray}
\frac{\partial}{\partial z}\hat{\Xi}_{\mathrm{I}}(z,t)&=&%
-2\epsilon\,\bar{\Xi}_2\,\hat{\mathcal T}_{\mathrm{I}}(z,t)%
\nonumber\\%
\frac{\partial}{\partial z}\hat{\Xi}_{\mathrm{I\!I\!I}}(z,t)&=&%
\phantom{+}2\epsilon\,\bar{\Xi}_2\,\hat{\mathcal T}_{\mathrm{I\!I\!I}}(z,t)%
\nonumber\\%
\frac{\partial}{\partial t}\hat{\mathcal T}_{\mathrm{I}}(z,t)&=&%
\phantom{+}\bar{c}_{13}\epsilon\,\bar{\mathcal F}_{z}\,%
\hat{\Xi}_{\mathrm{I}}(z,t)\,%
\nonumber\\%
\frac{\partial}{\partial t}\hat{\mathcal T}_{\mathrm{I\!I\!I}}(z,t)&=&%
-\bar{c}_{13}\epsilon\,\bar{\mathcal F}_{z}\,%
\hat{\Xi}_{\mathrm{I\!I\!I}}(z,t)\,%
\label{b.3}%
\end{eqnarray}
Then its solution can be found by the method of Laplace transform.

To show this we can define the Laplace images of the space-time
dependent Stokes components of the probe light and of the
collective alignment components of atoms
\begin{eqnarray}
\hat{\Xi}_{\mathrm i}(p,s)&=&\int_0^\infty\int_0^\infty\! dz\,dt\,%
e^{-pz-st}\,\hat{\Xi}_{\mathrm i}(z,t)%
\nonumber\\%
\hat{{\mathcal T}}_\mu(p,s)&=&\int_0^\infty\int_0^\infty\! dz\,dt\,%
e^{-pz-st}\,\hat{{\mathcal T}}_\mu(z,t)%
\label{b.4}%
\end{eqnarray}%
with ${\mathrm i}=\mathrm{I,I\!I\!I} $ and
$\mu=\mathrm{I,I\!I\!I}$. Then equations (\ref{b.3}) is
transformed to the system of algebraic equations, which in turn
can be straightforwardly solved similar to how it was done in
Appendix C of Ref.\cite{KMSJP}. The important parameter for this
procedure is the determinant of the system, which is given by
\begin{equation}
\Delta(p,s)\;=\;\left[sp+%
2\bar{c}_{13}\epsilon^2\bar{\Xi}_2\bar{\mathcal F}_z\right]^2%
\label{b.5}%
\end{equation}%
and its pole
\begin{eqnarray}
s&=&\frac{A}{p}%
\label{b.6}%
\end{eqnarray}
with $A=-2\bar{c}_{13}\epsilon^2\bar{\Xi}_2\bar{\mathcal F}_z$,
can be linked with the dispersion law for the spin polariton wave
propagating through the sample, see Eq.(\ref{3.1}) and discussion
around it.

Finally for $A<0$ one obtains the following solution for the
Stokes components of the light subsystem
\begin{widetext}
\begin{eqnarray}
\hat{\Xi}_{\mathrm{I}}(L,t)&=&%
\hat{\Xi}_{\mathrm{I}}^{\mathrm{in}}(t)-\int_0^t\!dt'%
\left[\frac{-AL}{t-t'}\right]^{1/2}%
J_1\left(2\left[-AL(t-t')\right]^{1/2}\right)\hat{\Xi}_{\mathrm{I}}^{\mathrm{in}}(t')%
\nonumber\\%
&&-2\epsilon \bar{\Xi}_2\int_0^L\!dz\,%
J_0\left(2\left[-A(L-z)t\right]^{1/2}\right)%
\hat{{\mathcal T}}_{\mathrm{I}}^{\mathrm{in}}(z)%
\nonumber\\%
\hat{\Xi}_{\mathrm{I\!I\!I}}(L,t)&=&%
\hat{\Xi}_{\mathrm{I\!I\!I}}^{\mathrm{in}}(t)-\int_0^t\!dt'%
\left[\frac{-AL}{t-t'}\right]^{1/2}%
J_1\left(2\left[-AL(t-t')\right]^{1/2}\right)\hat{\Xi}_{\mathrm{I\!I\!I}}^{\mathrm{in}}(t')%
\nonumber\\%
&&+2\epsilon \bar{\Xi}_2\int_0^L\!dz\,%
J_0\left(2\left[-A(L-z)t\right]^{1/2}\right)%
\hat{{\mathcal T}}_{\mathrm{I\!I\!I}}^{\mathrm{in}}(z)%
\label{b.7}%
\end{eqnarray}
and for the alignment components of the spin subsystem
\begin{eqnarray}
\hat{\mathcal T}_{\mathrm{I}}(z,T)&=&%
\hat{\mathcal T}_{\mathrm{I}}^{\mathrm{in}}(z)-\int_0^z\!dz'%
\left[\frac{-AT}{z-z'}\right]^{1/2}%
J_1\left(2\left[-AT(z-z')\right]^{1/2}\right)\hat{\mathcal T}_{\mathrm{I}}^{\mathrm{in}}(z')%
\nonumber\\%
&&+\bar{c}_{13}\epsilon \bar{\mathcal F}_z%
\int_0^T\!dt\,J_0\left(2\left[-A(T-t)z\right]^{1/2}\right)%
\hat{\Xi}_{\mathrm{I}}^{\mathrm{in}}(t)%
\nonumber\\%
\hat{\mathcal T}_{\mathrm{I\!I\!I}}(z,T)&=&%
\hat{\mathcal T}_{\mathrm{I\!I\!I}}^{\mathrm{in}}(z)-\int_0^z\!dz'%
\left[\frac{-AT}{z-z'}\right]^{1/2}%
J_1\left(2\left[-AT(z-z')\right]^{1/2}\right)\hat{\mathcal T}_{\mathrm{I\!I\!I}}^{\mathrm{in}}(z')%
\nonumber\\%
&&-\bar{c}_{13}\epsilon \bar{\mathcal F}_z%
\int_0^T\!dt\,J_0\left(2\left[-A(T-t)z\right]^{1/2}\right)%
\hat{\Xi}_{\mathrm{I\!I\!I}}^{\mathrm{in}}(t)%
\label{b.8}%
\end{eqnarray}
\end{widetext}
Here $L$ is the sample length and $T$ is the interaction time. The
solution performs an integral transform of the Heisenberg
operators for the Stokes components in the incident light
$\hat{\Xi}_{\mathrm i}^{\mathrm{in}}(t)$ and of initial
Schr\"{o}dinger operators for the alignment components of atomic
spins $\hat{\mathcal T}_{\mu}^{\mathrm{in}}(z)$. The kernels of
the transform are expressed by the cylindrical Bessel functions of
the zeroth $J_0(\ldots)$ and first $J_1(\ldots)$ order. The
original form of atomic and field operators can be recovered with
the orthogonal transformations reverse to (\ref{b.1}) and
(\ref{b.2}).

The derived solution preserves the following commutation relations
\begin{eqnarray}
\left[\hat{\Xi}_{\mathrm{I\!I\!I}}(z,t), \hat{\Xi}_{\mathrm{I}}(z,t')\right]&=&%
2i\,\delta(t-t')\bar{\Xi}_2%
\nonumber\\%
\left[\hat{\mathcal T}_{\mathrm{I\!I\!I}}(z,t), \hat{\mathcal T}_{\mathrm{I}}(z',t)\right]&=&%
-i\bar{c}_{13}\,\delta(z-z')\,\bar{\mathcal F}_{z}%
\label{b.9}%
\end{eqnarray}
The commutation relation for the Stokes components, given by the
first line, differs from the original commutation relation in form
(\ref{2.4}). That is direct consequence of our ignoring the
retardation effects. The argument of $\delta$-function, which
generally is $t-t'-(z-z')/c$, can be only approximately reproduced
in the assumptions we did. The solution also obeys the following
important cross-type commutation relations between the field and
atomic variables
\begin{eqnarray}
\lefteqn{\left[\hat{\Xi}_{\mathrm i}(z,t),\hat{\mathcal T}_{\mu}(z',t')\right]=%
-2i\bar{c}_{13}\epsilon\,g_{{\mathrm i}\mu}\bar{\Xi}_2\bar{\mathcal F}_{z}}%
\nonumber\\%
&&\times J_0\left(2\left[-A(z-z')(t-t')\right]^{1/2}\right)%
\left[\theta(t-t')-\theta(z'-z)\right]%
\nonumber\\%
\label{b.10}%
\end{eqnarray}
where the matrix $g_{{\mathrm i}\mu}$ (with ${\mathrm
i},\mu=\mathrm{I,I\!I\!I}$) is given by
\begin{equation}
g_{{\mathrm i}\mu}=%
\left(\begin{array}{cc}%
0,&%
1 \\%
1,&%
0%
\end{array}\right)
\label{b.11}
\end{equation}
These commutation relations clear indicate that the atomic and
field variables always commute before and after interaction.

For $A>0$ one obtains the following solution for the Stokes
components of the light subsystem
\begin{widetext}
\begin{eqnarray}
\hat{\Xi}_{\mathrm{I}}(L,t)&=&%
\hat{\Xi}_{\mathrm{I}}^{\mathrm{in}}(t)+\int_0^t\!dt'%
\left[\frac{AL}{t-t'}\right]^{1/2}%
I_1\left(2\left[AL(t-t')\right]^{1/2}\right)\hat{\Xi}_{\mathrm{I}}^{\mathrm{in}}(t')%
\nonumber\\%
&&-2\epsilon \bar{\Xi}_2\int_0^L\!dz\,%
I_0\left(2\left[A(L-z)t\right]^{1/2}\right)%
\hat{{\mathcal T}}_{\mathrm{I}}^{\mathrm{in}}(z)%
\nonumber\\%
\hat{\Xi}_{\mathrm{I\!I\!I}}(L,t)&=&%
\hat{\Xi}_{\mathrm{I\!I\!I}}^{\mathrm{in}}(t)+\int_0^t\!dt'%
\left[\frac{AL}{t-t'}\right]^{1/2}%
I_1\left(2\left[AL(t-t')\right]^{1/2}\right)\hat{\Xi}_{\mathrm{I\!I\!I}}^{\mathrm{in}}(t')%
\nonumber\\%
&&+2\epsilon \bar{\Xi}_2\int_0^L\!dz\,%
I_0\left(2\left[A(L-z)t\right]^{1/2}\right)%
\hat{{\mathcal T}}_{\mathrm{I\!I\!I}}^{\mathrm{in}}(z)%
\label{b.12}%
\end{eqnarray}
and for the alignment components of the spin subsystem
\begin{eqnarray}
\hat{\mathcal T}_{\mathrm{I}}(z,T)&=&%
\hat{\mathcal T}_{\mathrm{I}}^{\mathrm{in}}(z)+\int_0^z\!dz'%
\left[\frac{AT}{z-z'}\right]^{1/2}%
I_1\left(2\left[AT(z-z')\right]^{1/2}\right)\hat{\mathcal T}_{\mathrm{I}}^{\mathrm{in}}(z')%
\nonumber\\%
&&+\bar{c}_{13}\epsilon \bar{\mathcal F}_z%
\int_0^T\!dt\,I_0\left(2\left[A(T-t)z\right]^{1/2}\right)%
\hat{\Xi}_{\mathrm{I}}^{\mathrm{in}}(t)%
\nonumber\\%
\hat{\mathcal T}_{\mathrm{I\!I\!I}}(z,T)&=&%
\hat{\mathcal T}_{\mathrm{I\!I\!I}}^{\mathrm{in}}(z)+\int_0^z\!dz'%
\left[\frac{AT}{z-z'}\right]^{1/2}%
I_1\left(2\left[AT(z-z')\right]^{1/2}\right)\hat{\mathcal T}_{\mathrm{I\!I\!I}}^{\mathrm{in}}(z')%
\nonumber\\%
&&-\bar{c}_{13}\epsilon \bar{\mathcal F}_z%
\int_0^T\!dt\,I_0\left(2\left[A(T-t)z\right]^{1/2}\right)%
\hat{\Xi}_{\mathrm{I\!I\!I}}^{\mathrm{in}}(t)%
\label{b.13}%
\end{eqnarray}
\end{widetext}
Here the functions $I_0(\ldots)$ and $I_1(\ldots)$ are the
modified Bessel functions of the zeroth and first order. The
solution obeys the commutation relations (\ref{b.9}) and the
cross-type commutation relations are now given by
\begin{eqnarray}
\lefteqn{\left[\hat{\Xi}_{\mathrm i}(z,t),\hat{\mathcal T}_{\mu}(z',t')\right]=%
-2i\bar{c}_{13}\epsilon\,g_{{\mathrm i}\mu}\bar{\Xi}_2\bar{\mathcal F}_{z}}%
\nonumber\\%
&&\times I_0\left(2\left[A(z-z')(t-t')\right]^{1/2}\right)%
\left[\theta(t-t')-\theta(z'-z)\right]%
\nonumber\\%
\label{b.14}%
\end{eqnarray}

\end{document}